\documentclass[12pt,epsf,amssymb]{article}
\usepackage{color}
\usepackage{graphics}
\usepackage{epsfig}

\makeatletter

 \usepackage{verbatim}

\newcommand{\msbar}{\overline{\rm MS}}
\newcommand{\bea}{\begin{eqnarray}}
\newcommand{\eea}{\end{eqnarray}}
\newcommand{\beq}{\begin{equation}}
\newcommand{\eeq}{\end{equation}}
\newcommand{\gev}{{\rm GeV}}
\newcommand{\mev}{{\rm MeV}}

\newcommand{\pdir}{p\kern -5.2pt\raise 0.2ex\hbox {/}}
\newcommand{\vdir}{v\kern -5.75pt\raise 0.15ex\hbox {/}}
\newcommand{\kdir}{k\kern -5.75pt\raise 0.15ex\hbox {/}}
\newcommand{\epsdir}{\epsilon\kern -5.0pt\raise 0.15ex\hbox {/}}
\newcommand{\bvdir}{\bar{v}\kern -5.75pt\raise 0.15ex\hbox {/}}
\newcommand{\Ddir}{D\kern -7.75pt\raise 0.20ex\hbox {/}}
\newcommand{\ldir}{l\kern -5.0pt\raise 0.2ex\hbox{/}}
\newcommand{\varepsdir}{\varepsilon\kern -5.5pt\raise 0.15ex\hbox{/}}

\makeatother

\begin{document}
\thispagestyle{empty} 
{\par\raggedleft BUHEP-00-3\par}
{\par\raggedleft FTUV-IFIC-00-0216\par}
{\par\raggedleft LPT-Orsay/00-19 \par}
{\par\raggedleft RM3-TH/00-3\par}
{\par\raggedleft ROMA-1285/00\par}

\vskip 0.25cm

{\par\centering \textbf{\LARGE $B^0$--$\bar B^0$ Mixing and Decay Constants with}\\
\textbf{\LARGE the Non-Perturbatively Improved Action}\par}

{\par\centering \vskip 0.5 cm\par}

{\par\centering {\large D.~Becirevic, D.~Meloni and A.~Retico}\\
\vskip 0.25cm\par}

{\par\centering \textit{Dip.~di Fisica, Univ.~di Roma ``La 
Sapienza''
and}\\
\textit{INFN, Sezione di Roma, P.le A.~Moro 2, I-00185 Roma, 
Italy.}\\
\vskip 0.3cm\par}

{\par\centering {\large V.~Gim\'enez}\\
\vskip 0.25cm\par}

{\par\centering \textit{Dep.~de F\'isica Te\`orica and IFIC, Univ.~de 
Val\`encia,}\\
\textit{Dr.~Moliner 50, E-46100, Burjassot, Val\`encia, 
Spain}\textsf{\textit{.}}\\
\vskip 0.3cm\par}

{\par\centering {\large L.~Giusti}\\
\vskip 0.25cm\par}

{\par\centering \textit{Department of Physics, Boston University}\\
\textit{Boston, MA 02215 USA.}\\
\vskip 0.3cm\par}

{\par\centering {\large V.~Lubicz}\\
\vskip 0.25cm\par}

{\par\centering \textit{Dipartimento di Fisica, Universit\`a di Roma Tre
and INFN, Sezione di Roma Tre}\\
\textit{Via della Vasca Navale 84, I-00146 Rome, Italy}.\\
\vskip 0.3cm\par}

{\par\centering {\large G.~Martinelli}\\
\vskip 0.25cm\par}

{\par\centering \textit{Universit\'e de Paris Sud, LPT (B\^at.~210) }\\
\textit{Centre d'Orsay, 91405 Orsay-Cedex, France}.\\
\vskip 0.5cm\par}

\vskip 0.25cm
\hrule
\begin{abstract}
Several quantities relevant to phenomenological studies
of $B^0$--$\bar B^0$ mixing are computed  on the
lattice. Our main results are $f_{B_d}
\sqrt{\hat B_{B_d}}=206(28)(7)$ MeV, $\xi= f_{B_s}
\sqrt{\hat B_{B_s}}/f_{B_d}\sqrt{\hat B_{B_d}}=1.16(7)$. We also obtain
the related quantities $f_{B_s}\sqrt{\hat B_{B_s}}=237(18)(8)$ MeV, 
$f_{B_d}= 174(22)^{+7+4}_{-0-0}$ MeV, $f_{B_s}=
 204(15)^{+7+3}_{-0-0}$ MeV, $f_{B_s}/f_{B_d}=1.17(4)^{+0}_{-1} $, 
$f_{B_d}/f_{D_s}=0.74(5)$. After combining our results with the
experimental world average $\Delta m_d^{\rm (exp.)}$, we predict $\Delta m_s
=15.8(2.1)(3.3)\ ps^{-1}$. We have also computed the relevant  parameters
for $D^0$--$\bar D^0$ mixing which may be useful in some extensions of
the Standard Model. All the quantities were obtained from a quenched 
si\-mu\-la\-tion with a non-perturbatively improved Clover action at $\beta=6.2$,
corresponding to a lattice spacing $a^{-1}=2.7(1)$ GeV,
on a sample of 200 gauge-field configurations.  A discussion of the 
main systematic errors is also presented.
\end{abstract}
\hrule
\vskip 0.2cm
{\par\centering PACS: 13.75Lb,\ 11.15.Ha,\ 12.38.Gc. \par}
\vskip 0.2 cm 
\setcounter{page}{1}
\setcounter{footnote}{0}
\setcounter{equation}{0}
\section{Introduction}
\label{sec:intro}
Historically, the measurement  by the UA1 Collaboration
of  a large value for the neutral meson
mass-difference $\Delta m_d$ was the first  
indication that top quark mass had to be very heavy~\cite{UA1}.
This mass difference is  induced by $B_d^0$--$\bar B_d^0$ oscillations.
Since then, $B^0$--$\bar B^0$ mixing became  one of the most important
 ingredients of current analyses of
the unitarity triangle and of CP violation
in the Standard Model~\cite{lusignoli}--\cite{babar}
and beyond~\cite{barbieri}. 
\par In the Standard Model the mixing, induced by the box diagrams with an
internal top quark, is summarized by the next-to-leading order (NLO)
formula~\cite{buras,reviewbb}
\bea
\label{deltam}
\Delta m_q \ = 
\ {G_F^2\over 6 \pi^2} m_W^2 \ \eta_B S_0(x_t) 
\ \vert V_{tq} V_{tb}^*\vert^2 \ m_{B_q} f_{B_q}^2 \hat B_{B_q}\;\;\;
 (q=s,d)\;\; ,
\eea
where $S_0(x_t)$ is  the Inami-Lim function~\cite{inami}, $x_t=m_t^2/M^2_W$
($m_t$ is the $\msbar$ top mass
$m^{\msbar}_t(m_t^{\msbar})=165(5)$ GeV) and 
$\eta_B=0.55(1)$ is the perturbative QCD short-distance NLO correction.
The remaining factor, $f_{B_q}^2 \hat B_{B_q}$, encodes the information 
of non-perturbative QCD and this is what can be computed on the lattice. 
\par 
Traditionally, the $B$-parameter of 
the renormalized operator  is defined as  
\bea
\label{defB}
\langle  \bar B_{q} \vert   Q_q^{\Delta B=2} (\mu) \vert B_{q} 
\rangle  = \frac{8}{3}\ m_{B_q}^2 f_{B_q}^2 \ B_{B_q}(\mu)\, , 
\eea
where $Q_q^{\Delta B=2}= (\bar b \gamma_\mu (1-\gamma_5) q)
(\bar b \gamma_\mu (1-\gamma_5) q)$ and 
$\mu$ is the renormalization scale.
This definition stems from the vacuum saturation approximation (VSA)
in which $B_{B_q}=1$.
The renormalization group invariant, and scheme-independent,
 $B$ parameter $\hat B_{B_q}$ of eq.~(\ref{deltam}) is defined as
\bea
\label{bhat}
\hat B_{B_q} =  \alpha_s(\mu)^{-\gamma_0/2\beta_0}
 \left\{ 1 + {\alpha_s(\mu) \over 4 \pi} J\right\} \  B_{B_q}(\mu) \ ,
\eea
where $\gamma_0=4$ and $J$ depends on the scheme used 
for renormalizing  $Q_q^{\Delta B=2} (\mu)$ (see below).
\par Besides the $B^0$--$\bar B^0$ amplitude,
an important  quantity for phenomenological applications is given by the 
ratio
\bea \label{ratiom} \frac{\Delta m_s}{\Delta m_d} =\frac{\vert V_{ts}\vert^2}
{\vert V_{td}\vert^2} \frac{m_{B_s}}{m_{B_d}} \xi^2
 \ ,\eea
where $\xi= f_{B_s}\sqrt{ \hat B_{B_s} }/f_{B_d}\sqrt{ \hat B_{B_d} }$.
\vskip 0.4 cm
In  this paper, we have computed  the
hadronic parameters appearing in eqs.~(\ref{deltam}) and (\ref{ratiom}),
using a non-perturbatively improved action, and with operators renormalized 
on the lattice with the non-perturbative method of ref.~\cite{np},
as implemented in \cite{DELTAS=2,bibbia}.
Our main results are 
\bea  f_{B_d}
\sqrt{\hat B_{B_d}}&=& 206(28)(7) \, {\rm MeV}\ , \quad
\xi =  \frac{f_{B_s}
\sqrt{\hat B_{B_s}}}{f_{B_d}\sqrt{\hat B_{B_d}}}=1.16(7)\ , 
\nonumber \\
f_{B_s}\sqrt{\hat B_{B_s}}& =& 237(18)(8)\, {\rm MeV}\ , \quad
r_{sd} =  \xi^2 \left(\frac{m^2_{B_s}}{m^2_{B_d}}\right)=1.40(18)
 \ , \nonumber \\ 
f_{B_d} &= & 174(22)^{+7+4}_{-0-0} \,  {\rm MeV} \ , \quad
f_{B_s}=  204(15)^{+7+3}_{-0-0}\, {\rm MeV} \ , \label{all} \\
\frac{f_{B_s}}{f_{B_d}}&=& 1.17(4)^{+0}_{-1} \, , \quad
\frac{f_{B_d}}{f_{D_s}} =  0.74(5) \ . \nonumber  \eea 
From eq.~(\ref{deltam}) we write
\bea \Delta m_d \; [{ps}^{-1}]\
 =\ (0.153 \pm 0.010 )\ \vert V_{td} \vert^2 
\,  f_{B_d}^2 \hat B_{B_d} \,\; [{\rm MeV}^2] \ . \eea
We stress that what is really relevant for $B^0$--$\bar B^0$ mixing,
and can be directly computed on the lattice, is the physical
amplitude,  corresponding to $f^2_{B_d} \hat B_{B_d}$, and not the decay 
constant $f_{B_d}$ and  $\hat B_{B_d}$  separately. 
The value of $f_{B_d} \sqrt{\hat B_{B_d}}$ given in the first of 
eqs.~(\ref{all}) comes from the calculation of the amplitude,
and thus it includes the correlation
between the decay constant and the $B$ parameter. Its final error is then
smaller than that obtained by combining the results and errors obtained from
$f_{B_d}$ and $\hat B_{B_d}$ from different  calculations.

By taking  $\vert V_{td}\vert \simeq A \lambda^3 \sqrt{(1-\rho)^2 + \eta^2} 
=0.0080(5)$ (where $A$, $\rho$, $\eta$  and $\lambda$ are the Wolfenstein parameters) 
this gives 
\bea  
\Delta m_d = 0.42(12)(4) \, {\rm  ps}^{-1} \ , 
\eea
where the first error comes from the lattice uncertainty on $f_{B_d}
\sqrt{\hat B_{B_d}}$ and the second  from the error on $m_t$ and on
$\vert V_{td} \vert$. 
The most recent world average of experimental
results is~\cite{experiment}:
\bea
 \Delta m^{\rm (exp.)}_d = 0.473(16)\, { ps}^{-1} \ . \eea
To predict $\Delta m_s$ it is convenient to use
 the above experimental information and eq.~(\ref{ratiom}) written
as 
\bea \Delta m_s = \frac{\vert V_{ts}\vert^2} {\vert V_{td}\vert^2} \xi^2
\left( \frac{m_{B_s}}{m_{B_d}}
\, \Delta m_d \right)^{\rm (exp.)} \ . \eea
Using our result for $\xi^2$ and $\vert V_{ts}\vert^2/\vert V_{td}\vert^2\simeq 
1/[\lambda^2 ((1-\rho)^2 + \eta^2] = 24.4(5.0) $, we get
\bea \Delta m_s = 15.8 (2.1)(3.3) \, { ps}^{-1} \ , \eea 
 to  be compared with the experimental lower bound
\bea \Delta m_s >  12.4 \ ps^{-1} \ . \eea
In the above calculations, we have assumed the values of the relevant 
couplings, namely $\vert V_{td}\vert$ and $\vert V_{ts}\vert$, from the unitarity 
relations of the $V_{\rm CKM}$  matrix.
Obviously, from the experimental determinations of
$\Delta m_q$ and the hadronic matrix
elements  we can,  instead, constrain  $\vert V_{tq}\vert$. This is what is usually done
in the unitarity triangle analyses~\cite{lusignoli}--\cite{barbieri}.
For the sake of comparison, we also give the values of the decay constants and their
ratios from our previous studies of these quantities, obtained using the same improved
action, with a comparable statistics but on a larger lattice~\cite{heavy}
\bea  f_{B_d} &= & 173(13)^{+34}_{-2} {\rm MeV} \ , \quad
f_{B_s}= 196(11)^{+42}_{-0} {\rm MeV} \ , \\
\frac{f_{B_s}}{f_{B_d}}&=& 1.14(2)(1)\, , \quad
\frac{f_{B_d}}{f_{D_s}} =  0.72(2)^{+13}_{-0} \ . \nonumber\eea 
The last error in the figures above represents the uncertainty coming
from the extrapolation (linear vs quadratic) 
in the inverse meson mass   to the $B$ mesons. This error is not given in
eqs.~(\ref{all}),  because in the present  study only a linear extrapolations 
have been made for reasons discussed below.
\par Besides the quantities relevant to $B^0$--$\bar B^0$ mixing, we 
also give the corresponding quantities for $D^0$--$\bar D^0$ mixing. Although 
$D^0$--$\bar D^0$ mixing is expected to be well below the ex\-pe\-ri\-men\-tal limit in
the Standard Model, it may be enhanced in its extensions~\cite{nir}.
We have obtained
\bea  f_{D}
\sqrt{\hat B_{D}}&=& 230(14)^{+3}_{-8} \, {\rm MeV}\ , \quad
f_{D} =  207(11)^{+3+3}_{-0-0}\, {\rm MeV} \, ,  \\ 
f_{D_s}&= & 234(9)^{+3+2}_{-0-0}\, {\rm MeV}\, ,  \quad
\frac{f_{D_s}}{f_{D}}= 1.13(3)^{+0}_{-1} \ . \nonumber \eea 
Reviews of recent results of quantities considered in this paper can be found in
refs.~\cite{reviews}.
\par The remainder of this paper is as follows: in sec.~\ref{sec:lcdc} we give the
parameters of the lattice simulation, illustrate  the calibration of the
lattice spacing and  of the quark masses and describe the calculation of the
heavy meson spectrum and decay constants; in sec.~\ref{sec:4f} we
discuss the renormalization of the relevant operators, the calculation of their
matrix elements and the extrapolation to the physical points; in sec.~\ref{sec:phys}
we give the physical results and  discuss the statistical and systematic errors;
sec.~\ref{sec:con} contains a comparison of our  results with other  calculations
of the same quantities as well as our conclusions.
\section{Lattice Calibration and Decay Constants}
\label{sec:lcdc}
In this section we give some details about our lattice simulation  and the
calculation of the decay constants, which are a byproduct of this study.

\subsection{Lattice Setup} 
\label{subsec:ls} 
The results presented in this study have been  obtained  
on a $24^3\times 48$ lattice, using the 
non-perturbatively improved Clover action at $\beta=6.2$
with the Clover coefficient  $c_{_{SW}}=1.614$, as computed in ref.~\cite{csw}.
The statistical sample  consists of 200 independent gauge-field configurations.
Statistical errors have been  estimated by using the 
familiar jackknife procedure with 40 jacks,  each 
 obtained by decimating 5 configurations from the whole ensemble. 
The following values for the heavy- and light-quark hopping parameters,
 $\kappa_{Q}$  and  $\kappa_{q}$ respectively,  have been used:
\begin{itemize}
\item  0.1352 ($\kappa_{q_1}$); 0.1349 ($\kappa_{q_2}$); 0.1344 ($\kappa_{q_3}$),
\item 0.1250 ($\kappa_{Q_1}$); 0.1220 ($\kappa_{Q_2}$); 0.1190 ($\kappa_{Q_3}$) . 
\end{itemize}
Although we are not discussing  light mesons, it is important to mention 
some results which will be useful 
in the analysis of  heavy-light meson physics, for example 
for the extrapolation/interpolation in light-quark masses. More details on
the procedures used to calibrate the lattice spacing and fix the light-quark masses
can be found in previous publications of our group~\cite{heavy,light}.
\begin{itemize}
\item[$\circ$] The ratio of the masses of the light pseudoscalar (P) and vector (V)  mesons are:
\bea
{m_P\over m_V} \ =\ \{  0.597(22)_{q_1},\ 0.682(15)_{q_2},\ 0.761(8)_{q_3} \}\ .
\eea 
\item[$\circ$] We use the method of physical lattice planes~\cite{lpmethod} to fix 
the value of the inverse lattice spacing~\footnote{ As in previous publications,
we use  small-case letters for referring to  quantities in physical units 
({\it e.g.} $m_P$ in MeV), and capital letters for denoting the same
 quantities in lattice units ({\it e.g.} $M_P = m_P a$).}. This method consists in
 the following. First we fit the vector meson mass to the form
\bea
\label{lp}
M_V(m_q, m_q) \ =\ \alpha_0 + \alpha_1  \left( M_P(m_q, m_q)\right)^2 \ ,
\eea
obtaining $\alpha_0 = 0.286(16)$, and $\alpha_1 = 1.24(13)$. Then 
we fix  $m_V/m_P$ to the physical kaon ratio  ($m_{K^*}/m_{K}$) and compare 
 $M_{K^*}$ to the experimentally measured $m_{K^*}=894$~MeV. In this way we obtain
\bea
a^{-1} = 2.72(13)\ {\rm GeV}\;.
\eea
This is the value of the inverse lattice spacing which has been 
used throughout this study.
\item[$\circ$] For a generic physical quantity in the heavy-light meson sector,
${\cal F}(m_Q,m_q)$, the interpolation/extrapolation in the light quark
to the strange/up-down ($s/d$) mass is performed through the fit
\bea
\label{lp2}
{\cal F}(m_Q,m_q) = \alpha_0^Q + \alpha_1^Q  \left( M_P(m_q, m_q)\right)^2\,,
\eea
where $\alpha_{0,1}^Q$ are the fitting parameters. 
The light-quark mass corresponds either to  the $d$-quark,
when  we extrapolate to  $M_P(m_d,m_d)\equiv M_\pi  = 5.1(3)\cdot 10^{-2}$
(as obtained from the lattice-plane method from $m_{\rho}/m_{\pi}$), or
to the strange quark, when we  extrapolate to
$M_P(m_s,m_s)\equiv M_{\eta_{ss}} = 0.266(17)$
(as inferred from eq.~(\ref{lp}) by  fixing $M_V(m_s,m_s)$ to $m_\phi=1020$~MeV).
\end{itemize}
\subsection{ Heavy-light Decay Constants}
\label{subsec:dc}
We now discuss the  heavy-light meson decay constants, 
which are important  ingredients for  physical predictions related to 
$B^0$--$\bar B^0$ mixing.
Since this has been extensively discussed in the literature (see for 
example ref.~\cite{heavy}), we only recall here the essential steps.
\begin{itemize} 
\item
As usual, hadron masses are extracted by fitting   two-point 
correlation functions. For mesons, the standard form is
\bea
{\cal C}_{_{JJ}}(t) = \sum_{\vec x} \langle 0 \vert J({\vec x}, t) 
J^{\dagger}(0)\vert 0 \rangle\, \stackrel{t\gg 0}{\longrightarrow}\, {{\cal{Z}}_J \over  M_J}e^{- M_J T/2} {\rm cosh}\left[ M_J \left( {T\over 2} - t \right) \right] ,
\label{eq : meff}
\eea
where for $J(x)=P_5(x)=i\bar{Q}(x)\gamma_5 q(x)$, we choose $t\in [16,23]$,
whereas for $J(x)=V_i(x)=\bar{Q}(x)\gamma_i q(x)$, 
we choose $t\in [19,23]$. The time intervals are established in the 
standard way ({\it i.e.} after inspection of the corresponding effective masses).
The resulting masses of the pseudoscalar and vector heavy-light mesons, with the 
light quark interpolated/extrapolated to $s/d$ (see eq.~(\ref{lp2})), are listed 
in tab.~\ref{tab1}. We also checked that on the same interval ($t\in [16,23]$), 
the masses of pseudoscalar mesons extracted from the correlation function 
 ${\cal C}_{AP}(t)$ ($A_0(x)=\bar{Q}(x)\gamma_0 \gamma_5 q(x)$), 
are indeed the same.
\par In physical units, the masses directly accessed from our simulation, are:
\bea
&&m_{P_d} = \{ 1.75(8)\ {\rm GeV},\ 2.02(9)\ {\rm GeV},\ 2.26(11)\ {\rm GeV}\}\ \,,\cr
&&\hfill \cr
&&m_{P_s} = \{ 1.85(7)\ {\rm GeV},\ 2.11(9)\ {\rm GeV},\ 2.38(10)\ {\rm GeV}\}\ \,. 
\eea

\item
The pseudoscalar meson (P) decay constant is defined as
\bea
\langle 0\vert {\hat{A}}_0\vert P(\vec p = 0)\rangle = i {\hat{F}}_{P} M_{P}\ ,
\eea
where the `hat' symbol means that the appropriate renormalization constant $Z_A$ is taken into account. To determine the decay constant, one computes the ratio
\bea
{{\sum_{\vec x} \langle \hat{A}_0(\vec x,t) P_5(0) \rangle } \over {\sum_{\vec x} \langle P_5(\vec x,t) P_5(0) \rangle }}  &\simeq&  {\hat{F}}_P \, {M_P\over {\sqrt {\cal{Z}}_P}}\, {\rm  tanh} \left(M_P (\frac{T}{2} - t)\right) \ .
\eea
When working with improved Wilson fermions, the axial current is  improved 
by adding to $A_\mu$ the operator $\partial_\mu  P$ with a suitable
coefficient 
\bea
\langle 0\vert \hat A_0 \vert P \rangle =
Z_A \Bigl(   \langle 0  \vert   A_0\vert  P \rangle + c_A \langle 0 \vert a 
\partial_0 P_5\vert P \rangle \Bigr)
= i M_{P} (\hat F_{P}^{(0)} + c_A a \hat F_{P}^{(1)})\,,
\eea
where the constant, $c_A=-0.037$, was computed in ref.~\cite{alpha}. 
It is  actually  easy to see that 
\beq
a F_{P}^{(1)} =  {{\sqrt {{\cal{Z}}_{P}}}\over M_{P}}  {\rm  sinh} (M_{P})\ .
\eeq
The size of the correcting term 
$c_A a \hat F_{P}^{(1)}/\hat F_{P}^{(0)}$ for our data is in the range $(5\div 7)\%$.
\item 
As far as the normalization constant is concerned,  its improved value is given by
(choice 1.)
\bea
\label{noklm}
  Z_A (m_Q, m_q) =Z_A(0) \, ( 1 + b_A   a \bar m )\ , \eea
where $\bar m=(m_Q + m_q)/2$. In the above equation, we have taken
 \bea a m_{Q,q} = {1\over 2}\left( {1\over \kappa_{Q,q}} - 
 {1\over \kappa_c} \right) \ .\eea 
The value of the critical hopping parameter, $\kappa_c = 
0.13580(1)$, is obtained from the linear fit $M_P^2 \simeq m_q \to 0$, 
whereas the value of improvement coefficient 
$b_A = 1.24$ in (\ref{klm}) is taken from boosted perturbation theory~\cite{bA}~\footnote{ 
By working with three light quark species only, we were unable to make a 
quadratic fit in the quark masses as done in ref.~\cite{light}.} For a recent non-perturbative 
determination of $b_A$, see ref.~\cite{lanl}.
. 
\par In our numerical estimates we have also 
used  \bea \label{klm}
Z_A (m_Q, m_q) =Z_A(0) \, \left[ {\sqrt{\ 1 + a m_Q}\,\sqrt{\ 1 + a m_q}\over 1
\,+\,a\bar m}\right]\ \left( 1\ +\  b_A\ a\bar m \right)\;,
\eea
This   equation  (choice 2.)  differs from the choice 1.  in that 
it contains  higher order tree-level mass corrections through the 
 so-called KLM factor~\cite{klm}.
As for the value of $Z_A\equiv Z_A(0)$ in the chiral limit, 
we take  the nonperturbative determination $Z_A(0)=0.80$
from ref~\cite{vittorio}. 
This value has been also non-perturbatively  computed  in refs.~\cite{lanl} 
($Z_A=0.82$) and \cite{ZA1} ($Z_A(0)=0.81$),  using different 
approaches. 
The  values of the lattice decay constants $\hat F_P$ for different
heavy-quark masses are  listed in the third column of tab.~\ref{tab1}. 
\end{itemize}
\begin{table}[h!]
\begin{center}
\begin{tabular}{|c|c|c|c|} 
  \hline
{\phantom{\huge{l}}}\raisebox{-.2cm}{\phantom{\Huge{j}}}
{  ``Flavor" content ($Q - q$)}& { ${\rm M_{P}}$} & { $\hat{F}_{\rm P}$} & ${\rm M_{V}}$  \\ \hline \hline
{\phantom{\Large{l}}}\raisebox{.2cm}{\phantom{\Large{j}}}

{\phantom{\Large{l}}}\raisebox{+.2cm}{\phantom{\Large{j}}}
{ \hspace{-1mm}$Q_3 - s$\hspace{1mm}} & {\sf 0.866(4)} & {\sf 0.0884(21)} & {\sf 0.898(4)}  \\    
{\phantom{\Large{l}}}\raisebox{-.2cm}{\phantom{\Large{j}}}
{\, \ $Q_3 - d$}  & {\sf 0.832(5)} & {\sf 0.0781(35)} & {\sf 0.869(7)}  \\ \hline   
{\phantom{\Large{l}}}\raisebox{.2cm}{\phantom{\Large{j}}}
{\phantom{\Large{l}}}\raisebox{+.2cm}{\phantom{\Large{j}}}
{ \hspace{-1mm}$Q_2 - s$\hspace{1mm}}  &{\sf  0.777(4)} & {\sf 0.0858(20)} & {\sf 0.812(4)} \\   
{\phantom{\Large{l}}}\raisebox{-.2cm}{\phantom{\Large{j}}}
{\, \ $Q_2 - d$}  & {\sf 0.742(3)} & {\sf 0.0761(29)} & {\sf 0.782(6)} \\ \hline  
{\phantom{\Large{l}}}\raisebox{.2cm}{\phantom{\Large{j}}}
{ \hspace{2mm}$Q_1 - s$\hspace{1mm}}  &{\sf 0.681(5)} & {\sf 0.0853(19)} & {\sf 0.721(5)} \\ 
{\phantom{\Large{l}}}\raisebox{-.2cm}{\phantom{\Large{j}}}
{\, $Q_1 - d $}  & {\sf 0.643(3)} & {\sf 0.0755(24)} & {\sf 0.690(5)}  \\ \hline  
\end{tabular}
\vspace*{.8cm}
\caption{\label{tab1}{\sl Mass spectrum of 
heavy-light pseudoscalar and vector mesons, with the light-quark masses
 extrapolated/interpolated to the $d/s$ quark, using eq.~{\rm (\ref{lp2})}.
 The meson decay constants are also given. All quantities are in lattice units.}}
\end{center}
\end{table}
To estimate the values of the physical quantities related to $B$-mesons 
($m_{B_d} = 5.28$~GeV, and $m_{B_s} = 5.38$~GeV), 
the extrapolations to the physical heavy-quark masses
necessarily rely on the heavy-quark symmetry.
 Accordingly,   the decay constants scale as
\bea
f_P =  {\Phi(m_P)\over \sqrt{m_P}} \left( 1 +{\Phi^\prime(m_P)\over \Phi(m_P)}{1 \over m_P}   + \dots \right) .
\eea
Apart from an overall (mild) logarithmic  correction, 
the coefficients $\Phi(m_P)$ and $\Phi^\prime(m_P)$, are  non-perturbative
quantities which we extract from a fit to the  lattice data. Since we work with 
three heavy quarks (three values of $\kappa_Q$), we have
only  included the leading $1/m_P$ correction by fitting the decay constant to
the expression
\bea
{\hat F}_P \sqrt{M_P} =  \Phi_0 \left( 1 + {\Phi_1 \over M_P} \right) .
\label{scaling}
\eea
In tab.~\ref{tab2}, we give  the physical results, 
and  the values of parameters $\Phi_{0,1}$, in physical units.
\begin{table}
\begin{center}
\begin{tabular}{|c|c|c|} 
\hline
{\phantom{\Large{l}}}\raisebox{+.2cm}{\phantom{\Large{j}}}
{ \hspace*{-4mm}$f_{ D_d}$}  & $f_{ D_s}$  & $f_{ D_s}/f_{D_d}$ \\    
{\phantom{\Large{l}}}\raisebox{-.5cm}{\phantom{\Large{j}}}
 $\mathsf 207(11)^{+3\ +3}_{-0\ -0}\ MeV$ &  $\mathsf 234(9)^{+3\ +2}_{-0\ -0}\ MeV$ & $\; \; \mathsf 1.13(3)^{+0\ +0}_{-0\ -1} \; \; $
   \\ \hline  
{\phantom{\Large{l}}}\raisebox{+.2cm}{\phantom{\Large{j}}}
{ \hspace*{-4mm}$f_{ B_d}$}  & $f_{ B_s}$  & $f_{ B_s}/f_{ B_d}$ \\    
{\phantom{\Large{l}}}\raisebox{-.4cm}{\phantom{\Large{j}}}
 $\mathsf 174(22)^{+7\ +4}_{-0\ -0}\ MeV$ &  $\mathsf 204(15)^{+7\ +3}_{-0\ -0}\ MeV$ & $\mathsf 1.17(4)^{+0\ +0}_{-0\ -1}$ \\ \hline  
 \multicolumn{3}{c}{ }   \\  
\end{tabular}
\vspace*{4mm}\\
\begin{tabular}{|c|c|} 
\hline
{\phantom{\Large{l}}}\raisebox{+.2cm}{\phantom{\Large{j}}}
{ \hspace*{-4mm}$\Phi_0^{(q=d)}$}  & $\Phi_1^{(q=d)}$  \\    
{\phantom{\Large{l}}}\raisebox{-.2cm}{\phantom{\Large{j}}}
 $0.46(7)\ {\rm GeV}^{3/2}$ &  $-0.73(13)\ {\rm GeV}$  \\ \hline
{\phantom{\Large{l}}}\raisebox{+.2cm}{\phantom{\Large{j}}}
  $\Phi_0^{(q=s)}$  & $\Phi_1^{(q=s)}$  \\    
{\phantom{\Large{l}}}\raisebox{-.2cm}{\phantom{\Large{j}}}
  $0.56(5)\ {\rm GeV}^{3/2}$ &  $-0.81(7)\ {\rm GeV}$   \\ \hline
\end{tabular}
\vspace*{.8cm}
\caption{\label{tab2}
{\sl Summary of  physical results for the decay constants. The parameters of 
eq.~{\rm (\ref{scaling})} are also listed. Our value for
$\Phi_1$ agrees well with  QCD sum rule estimates, namely
$\Phi_1=-0.9(2) \ {\rm GeV}$~{\rm \cite{braun}}.}}
\end{center}
\end{table}
The statistical errors are given in parentheses. The systematic errors 
require some explanation: the first one is obtained from the 
differences between    results  obtained by using $Z_A$ with 
choice 1. or 2; the second  is obtained by comparing the central values of the 
procedure described above to those  extracted from 
the study  of the ratio  $\hat F_P/M_V$. 
\par We have not included in the systematic errors: $i)$ the uncertainty due to 
quenching.   Pioneering attempts for estimating quenching  
errors~\cite{bernard}--\cite{cppacs} show 
an increase of about  $20\div 30$~MeV of the value of the decay constants.
$ii)$ the error due to 
the truncation of the  $1/m_P$ expansion in eq.~(\ref{scaling}).
The latter is usually 
estimated by considering also a quadratic term ($\propto 1/m_P^2$). 
With only three heavy-quark masses, we have
not done such a  fit.
From our previous experience~\cite{heavy}, we know that this error has a
negligible effect on the value of the
$D$-meson decay constants, whereas for the $B$ meson  it represents a
major source of systematics ($\sim 30$ MeV).
\par  Finally, we present an estimate of  $f_{B_d}$ obtained using
the experimental measurement of $f_{D_s} =  241 (32)$~MeV~\cite{roudeau}. 
With the ratio $f_B/f_{D_s}$ obtained in our simulation, table \ref{tab2}, we get
\bea
&&\left( f_{B_d} \over f_{D_s}\right)^{\rm (latt.)} = 0.74(4)^{+2 \ +2}_{-0 \ -0}=0.74 \pm 0.05 \, \cr
&& \hfill \cr 
\Longrightarrow && f_B = 184 \pm 24 {\rm (exp.)} \pm 12 {\rm (theo.)}\ {\rm
MeV}
\eea
\section{Matrix Elements of the Renormalized 
 Operators}
\label{sec:4f}
In this section, we discuss the renormalization of the relevant four-fermion
operators and describe the calculation of the matrix elements introduced
in eq.~(\ref{defB}). Since we are interested to the  $B^0$--$\bar B^0$ 
mixing amplitude,  only the  parity-even part of the operator,
$Q= (\bar b \gamma_\mu  q)
(\bar b \gamma_\mu  q) + (\bar b \gamma_\mu \gamma_5 q)
(\bar b \gamma_\mu \gamma_5 q)$, will be discussed in the following.
\subsection{Renormalization of the Four-fermion Operators}
\label{subsec:rffo}
On the lattice, the renormalized operator $ Q(\mu)$ 
takes the form~\cite{marti4f}
\bea
\label{subtr}
Q(\mu) = Z (\mu, g_0^2)
\ \left( O_1 + \sum_{i=2}^5 \Delta_i(g_0^2)  O_i \right)\ ,
\eea
where the $O_i$ denotes a bare operator and  we choose to 
 work in the following  parity-even basis 
\bea 
\label{basis}
 O_1 &=& \bar b \gamma_\mu  q \, 
\bar b \gamma_\mu  q + \bar b \gamma_\mu  \gamma_{5}  q \, 
\bar b \gamma_\mu \gamma_{5}  q \nonumber \\
 O_2 &=& \bar b \gamma_\mu  q \, 
\bar b \gamma_\mu  q - \bar b \gamma_\mu  \gamma_{5}  q \, 
\bar b \gamma_\mu \gamma_{5}  q \nonumber \\
 O_3 &=& \bar b   q \, 
\bar b  q - \bar b \gamma_{5}  q \, 
\bar b  \gamma_{5}  q  \label{ops} \\
 O_4 &=& \bar b  q \, 
\bar b q + \bar b   \gamma_{5}  q \, 
\bar b  \gamma_{5}  q \nonumber \\
 O_5 &=& \bar b \sigma_{\mu\nu}  q \, 
\bar b \sigma_{\mu\nu}  q  \;. \nonumber 
\eea 
To obtain the renormalized operator, two steps are then necessary.
First, one has to subtract 
operators of the same dimension, which (on the lattice) mix with 
$O_1$ (this is the consequence of the presence of 
the Wilson term in the fermion action, which explicitly breaks chiral 
symmetry). This means that one has to compute the subtraction constants
$\Delta_i(g_0^2)$. After an appropriate subtraction of the operators $O_i$, with
$i \neq 1$, the   matrix elements  still need  (to be finite) an
 overall renormalization constant  provided by $Z (\mu, g_0^2)$. A detailed
 discussion of the mixing matrix can be found in ref.~\cite{bibbia}.
In this work, we use the method of refs.~\cite{DELTAS=2,bibbia} to calculate the 
matching and the renormalization constants non-perturbatively. The computation, 
at three different values of the scale ($\mu a = 0.71$, $1.00$, $1.42$), 
is performed in the RI-MOM renormalization scheme,  in Landau gauge. 
The results are listed in table~\ref{tabZ}.
\begin{table} 
\begin{center} 
\begin{tabular}{|c|c|c|c|c|c|} 
\hline
{\phantom{\Large{l}}}\raisebox{+.2cm}{\phantom{\Large{j}}}
{ Scale $a \mu$}  & $Z(\mu , g_0^2)$   & $\Delta_2$ & $\Delta_3$ & $\Delta_4$ & $\Delta_5$ \\   \hline \hline {\phantom{\Large{l}}}\raisebox{+.2cm}{\phantom{\Large{j}}}
 0.71 &  0.630(11) & -0.052(3)& -0.016(2)& 0.014(2) &0.001(1) \\ \hline
{\phantom{\Large{l}}}\raisebox{-.2cm}{\phantom{\Large{j}}}
 1.01 &  0.583(7) & -0.055(2)& -0.017(2)& 0.016(2) &0.002(2)    \\ \hline
{\phantom{\Large{l}}}\raisebox{-.2cm}{\phantom{\Large{j}}}
 1.42 &  0.601(8) & -0.069(2)& -0.022(1)& 0.016(1) &0.007(2)   \\ \hline
\end{tabular}
\vspace*{.8cm}
\caption{\label{tabZ}
{\sl Summary of the results for renormalization and subtraction constants
(eq.~(\ref{subtr})),
evaluated non-perturbatively  at $\beta=6.2$ at three different scales $\mu$.  
$Z(\mu , g_0^2)$  is the renormalization constant in the RI-MOM scheme.}}
\end{center}
\end{table}
For sufficiently large $\mu$, we can use the available NLO anomalous dimension of 
the operator
$\langle   Q(\mu) \rangle^{\rm RI-MOM}$~\cite{ciuchini}, to define the 
renormalization group invariant (RGI) operator:
\bea
\label{anomalous}
\langle \, \hat  Q  \, \rangle = 
w^{-1}(\mu) \langle  Q(\mu)\rangle \equiv \alpha_s(\mu)^{-\gamma_0/2\beta_0} 
\left\{ 1 + {\alpha_s(\mu) \over 4 \pi} J\right\} \ \langle  Q (\mu)\rangle 
\ . \eea
where
\bea
\gamma_0 = 4\ ; && J_{\rm RI-MOM} \ = \ -\ {\; 17 397 -2070 n_f + 104 n_f^2\; 
 \over 6 ( 33 - 2 n_f)^2}\ +\ 8 \log 2 \, . \eea
We also give  the NLO part relevant 
 in the $\msbar$ scheme~\cite{buras}
\bea 
J_{\msbar }\ =\  {\; 13095 -1626 n_f + 8 n_f^2\;  \over 6 ( 33 - 2 n_f)^2}\; .
\eea
By using $\alpha_s(m_Z)=0.118$, and setting  $n_f=4$,  we obtain
\bea
\mu &=& \{ 1.9\ \gev,\; 2.73\ \gev,\; 3.85\ \gev \} \,,\cr
&& \cr
w(\mu)^{-1}_{\rm RI-MOM} &=& \{ \;  \; 1.408\; ,\hspace*{7mm} 1.453\; ,\hspace*{7mm}  1.489\; \; \;  \} \,.
\eea
In quenched calculations, there is always the embarrassing question whether
to use the physical value of $\alpha_s$, with the complete formulae for
the anomalous dimensions and  the $\beta$-function, or a ``quenched value''
of
the coupling constant (which suffers from intrinsic ambiguities), together with
anomalous dimensions and the $\beta$-function computed for $n_f=0$.
The present case is particularly lucky, since by computing $w(\mu)$
in the unquenched ($n_f=4$)  and quenched cases
 (with $\Lambda^{n_f=0}_{\rm QCD}=250$ MeV), one finds  values
of $\hat B_{B_d}$ which differ  by less than $2 \%$.

We give in passing,  $w(m_b)^{-1}_{\msbar} = 1.477$, which will be used 
later on for a comparison   with  the results of other authors. 
In the perturbative calculation of $w(m_b)^{-1}_{\msbar}$, we have used  
$m_b = 4.25$~GeV~\cite{guido}. 
\subsection{Matrix elements} 
We have computed the relevant three-point correlation functions:
\bea
{\cal C}^{(3)}_Q (\vec p, \vec q ; t_{P_1}, t_{P_2}) = 
 \int d\vec x d\vec y   \ e^{ i (\vec p \cdot \vec y -  
 \vec q\cdot \vec x )}\langle 0\vert P_5 (\vec x, t_{P_2}) 
  Q(\vec 0, 0)  P_5^\dagger (\vec y, t_{P_1}) \vert 0\rangle \ .
\eea
When the sources   are sufficiently 
separated and  far away from the origin, the lowest pseudoscalar mesons are 
isolated and one has
\bea
\label{3pts}
{\cal C}^{(3)}_Q (\vec p, \vec q ; t_{P_1}, t_{P_2})
  \stackrel{t_{P_1}, t_{P_2}\gg 0}{\longrightarrow}\, 
  {\sqrt{\cal{Z}}_P \over 2 E_{P}(\vec q)}e^{- E_P(\vec q) t_{P_1} }\  
  \langle \bar P (\vec q)\vert Q \vert P (\vec p)\rangle\  {\sqrt{\cal{Z}}_P 
  \over 2 E_{P}(\vec p)} e^{- E_P(\vec p) t_{P_2} } ,
\eea
where ${\cal{Z}}_P\equiv \langle 0\vert P_5 \vert P\rangle$. 
In order to eliminate irrelevant factors and to extract the $B$ pa\-ra\-me\-ter
we form the following ratio
\bea
\label{ratio}
{\cal R} (\vec p, \vec q; \mu) \ =\ {3\over 8}\ { {\cal C}^{(3)}_{ Q} (\vec p, \vec q ; t_{P_1}, t_{P_2}; \mu) 
\over \ {\cal C}_{AP}^{(2)} (\vec p ; t_{P_2}) \ {\cal C}_{AP}^{(2)} (\vec q ; t_{P_1})}\,,
\eea
where $\mu$ indicates the scale dependence. 
The two point correlation functions are
both multiplied by the axial-current renormalization constant 
in the chiral limit $Z_A(0)$.  We have not used the improved 
renormalization constant $Z_A(m_Q, m_q)$ since we have not attempted to improve 
the operator $Q$.  This is  beyond the scope of  the present study and 
would demand the inclusion of many
 operators of dimension seven, whose coefficients are presently unknown. 
Our hope is  that a part of the uncertainties of  ${\cal O}(\bar m a)$ 
 cancel in the ratio~(\ref{ratio}).
This has been  computed  at  $|\vec p| = |\vec q| 
\equiv 0$, corresponding to 
\bea
{\cal  R} (\mu)\  \to \ {3\over 8}\ {\langle \bar P \vert  Q(\mu)  \vert P \rangle \over
\ \vert \langle 0\vert \hat A_0 \vert P\rangle \vert^2}\,  \equiv\, B_{P_q}(\mu)\, .
\eea
In this way we obtain the main results of this work.
We also considered reasonably low momentum injections to the sources
but the additional noise makes these data not useful in practice.
For this reason these data will be ignored in the following. 
\par
\begin{figure}
\begin{center}
\begin{tabular}{@{\hspace{-0.7cm}}c c c}
&\epsfxsize12.0cm\epsffile{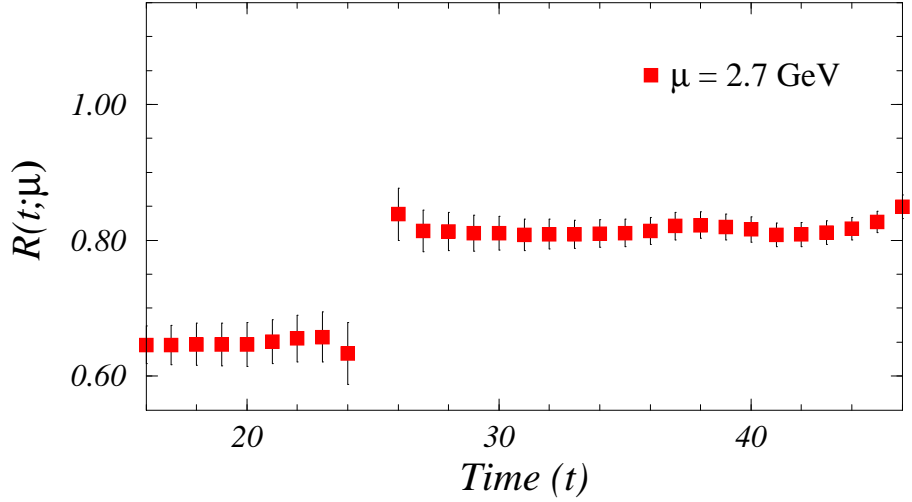} &  \\
\end{tabular}
\caption{\label{fig1}{\sl Ratio ${\cal R}(\mu)$ 
at $\mu a \simeq 1.42$, which  corresponds to $\mu \simeq 3.8$ GeV. 
The values of the heavy and light hopping parameters
are $\kappa_q=0.1349$ and $\kappa_Q = 0.1220$, respectively.}}
\end{center}
\end{figure}

 
 In order to show the quality of the signal,
in fig.~\ref{fig1}  we show the ratio ${\cal   R}(\mu)$  defined in eq.~(\ref{ratio}) 
at $\mu=3.8$ GeV ($\mu a =1.42$), for a specific combination of the heavy 
and light hopping parameters ($\kappa_Q = \kappa_{Q_2}=0.1220$ and 
$\kappa_q=\kappa_{q_2}=0.1349$). The ratio is plotted as a function of the 
time-distance of one of the two sources. The other has 
been fixed to $t_{P_1}=16$.  We also performed the simulation by fixing 
 $t_{P_1}=12$ and $t_{P_1}=20$. From the study of the two-point correlation functions, 
one isolates the lightest pseudoscalar state around $t=16$. That makes $t_{P_1}=12$
rather small for our purposes. For $t_{P_1}=20$, the sources are not far 
enough from each other because of the periodic boundary conditions,
and this  may spoil the signal.   
\begin{figure}
\begin{center}
\begin{tabular}{c c c}
&\epsfxsize8.0cm\epsffile{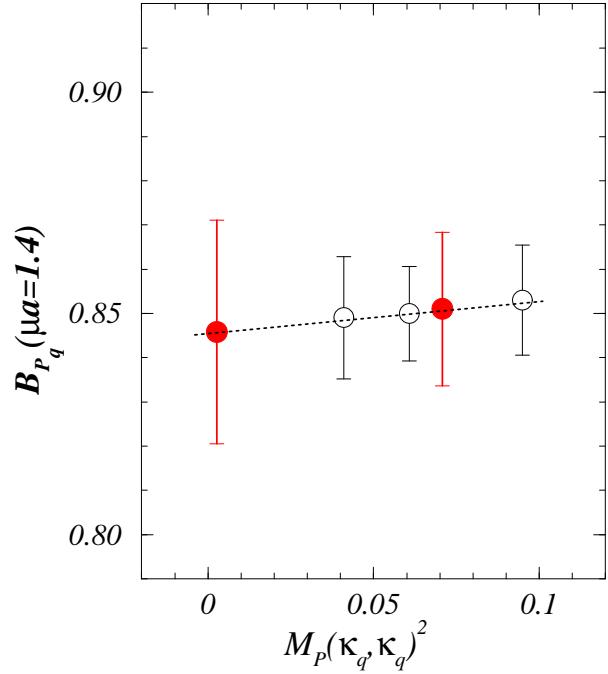} &    \\
\end{tabular}
\caption{\label{lp_otto}{\sl Fit of  ${\cal R} (\mu)
\equiv B_{P_q}(\mu)$ in the light-quark mass according to eq.{\rm (\ref{lp2})}.
 Empty symbols denote  the data obtained from our simulation ; filled symbols 
 represent $B_{P_s}(\mu)$ and 
$B_{P_d}(\mu)$, obtained after interpolation to the strange and extrapolation 
to the down  quark mass. The heavy quark corresponds  to $\kappa_{Q_2} = 0.1220$.}}
\end{center}
\end{figure}
After inspection of the ratios ${\cal R} (\mu)$ (for 
each combination of the heavy and light quarks), a stability plateau is 
found in the range $28 \leq t\leq 33$. From this plateau
we extract the values of ${\cal R} (\mu)$, that is the $B$ parameters,
for each combination of quark masses.
At this point,  we extrapolate the light quark to the $s$- and $d$-quark mass by 
fitting the $B$ parameter to eq.~(\ref{lp2}). This is illustrated 
in fig.~\ref{lp_otto} for $\kappa_{Q_2}= 0.1220$ at  $\mu=3.8$ GeV. In tab.~\ref{rrr1}, 
we give a detailed list of results for  the RI-MOM  $B_{P_q}(\mu)$, 
as well as  for the RGI $\hat B_{P_q}$ (see eq.~(\ref{anomalous})). 

\begin{table}[h!]
\vspace*{1.cm}
\begin{center}
\hspace*{-.7cm}
\begin{tabular}{||c|c|c|c|c|c|c||} 
  \hline\hline
{\phantom{\huge{l}}}\raisebox{-.2cm}{\phantom{\Huge{j}}}
{  Scale ($\mu$) }&  \multicolumn{2}{c|}{ $1.9$~GeV}  & 
 \multicolumn{2}{c|}{ $2.7$~GeV}  & 
 \multicolumn{2}{c||}{ $3.8$~GeV} \\  \hline 
 {\phantom{\Huge{l}}}\raisebox{-.1cm}{\phantom{\Huge{j}}}
{ Light quark  }&  $q =s$&  $q =d$ &  $q =s$&  $q =d$ &  $q =s$&  $q =d$ 
 \\  \hline \hline
{\phantom{\Huge{l}}}\raisebox{.17cm}{\phantom{\Huge{j}}}
$B_{P_q}(\mu;\kappa_Q=0.1250)$ & 0.853(16) & 0.842(23) &
0.789(15) & 0.779(21) & 0.831(15) & 0.821(21)  \\ 
{\phantom{\Huge{l}}}\raisebox{-.25cm}{\phantom{\Huge{j}}}
${\hat B}_{P_q}(\kappa_Q=0.1250)$ & {\sf 1.201(23)} & {\sf 1.186(32)} & 
{\sf 1.148(21)} & {\sf 1.132(30)} & {\sf 1.237(21)} & 
{\sf 1.222(29)} \\   \hline 
{\phantom{\Huge{l}}}\raisebox{+.17cm}{\phantom{\Huge{j}}}
$B_{P_q}(\mu;\kappa_Q=0.1220)$ & 0.875(19) & 0.867(27) &
0.810(18) & 0.803(25) & 0.851(18) & 0.846(25)  \\ 
{\phantom{\Huge{l}}}\raisebox{-.25cm}{\phantom{\Huge{j}}}
${\hat B}_{P_q}(\kappa_Q=0.1220)$ & {\sf 1.233(26)} & {\sf 1.222(39)} & 
{\sf 1.178(25)} & {\sf 1.167(37)} & {\sf 1.264(25)} & 
{\sf 1.259(38)} \\    \hline 
{\phantom{\Huge{l}}}\raisebox{.17cm}{\phantom{\Huge{j}}}
$B_{P_q}(\mu;\kappa_Q=0.1190)$ & 0.880(18) & 0.879(29) &
0.815(17) & 0.814(27) & 0.855(16) & 0.854(27)  \\ 
{\phantom{\Huge{l}}}\raisebox{-.25cm}{\phantom{\Huge{j}}}
${\hat B}_{P_q}(\kappa_Q=0.1190)$ & {\sf 1.239(25)} & {\sf 1.238(41)} & 
{\sf 1.184(23)} & {\sf 1.182(39)} & {\sf 1.273(24)} & 
{\sf 1.272(39)} \\   \hline\hline 
\end{tabular}
\vspace*{.25cm}
\caption{\label{rrr1}
{\sl $B_{P_q}$ for the three values of $\kappa_Q$
and with the light-quark mass extrapolated/interpolated to the $d/s$ quarks using 
eq.~{\rm (\ref{lp2})}. 
The results are given for both  the RI-MOM scheme and the RGI case.}}
\end{center}
\vspace*{.5cm}
\end{table}
\par
\begin{figure}[h!]
\vspace*{.8cm}
\begin{center}
\begin{tabular}{c c c}
&\epsfxsize11.7cm\epsffile{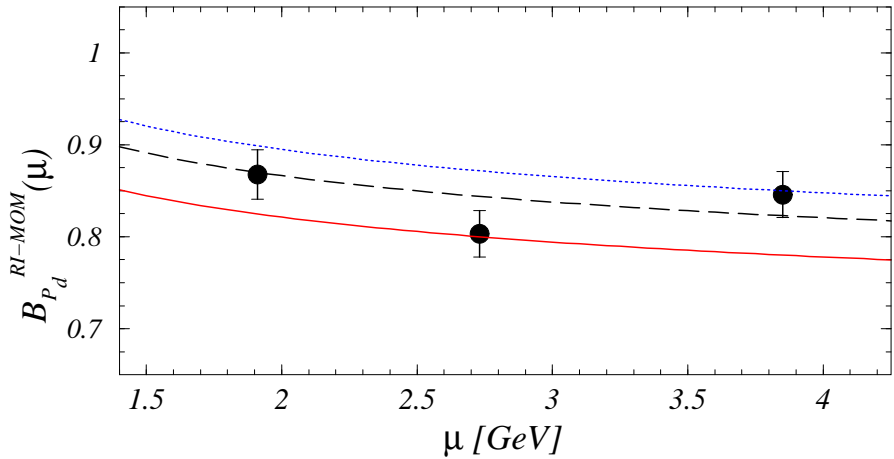} &    \\
\end{tabular}

\caption{\label{fig_evol}{\sl NLO evolution of $B_{P_d}^{\rm RI-MOM} (\mu)$ 
($\kappa_Q=0.1220$). The dashed, solid and dotted lines are obtained by starting the evolution from the lattice result (filled circles) at $\mu a
 = 0.71, 1.01, 1.42$, respectively.}}
\end{center}
\end{figure}

 It is interesting to check whether the scale 
dependence of the extracted $B_{P_q}(\mu)$ is well described by the 
NLO anomalous 
dimension (eq.(~\ref{anomalous})). In fig.~\ref{fig_evol}, we plot the 
evolution of  $B_{P_q}^{\rm RI-MOM}(\mu)$. The three  curves (dashed, full and dotted) have 
been normalized to the $B$ parameter  at the scale 
 $\mu a =0.71$, $1.01$ and $1.42$ respectively, as given in table~\ref{rrr1}.
We note that the point at $\mu a=1.01$ is  sensibly lower than what 
expected on the basis of the perturbative evolution. This   effect is a consequence of
a fluctuation of the value of $Z(\mu)$ at this value of the scale 
as shown in table \ref{tabZ}.  We checked that the fluctuation is enhanced by the
extrapolation of $Z(\mu)$ to the chiral limit and consider it as part of the
uncertainty in the determination of the matrix element. 
\noindent
The central values of ${\hat B}_{P_q}$ quoted below
correspond to the result obtained converting
the RI-MOM $B$ parameters  at $\mu a = 1.42$ to the RGI one.
The difference  with the results  obtained by using 
the other two scales, namely  $\mu a = 1$ and  $\mu a = 0.7$,  is
included in the   estimate of the systematic uncertainty.

\newpage
\section{Physical results}
\label{sec:phys}
\begin{figure}[h!]
\begin{center}
\begin{tabular}{c c c}
&\epsfxsize14.7cm\epsffile{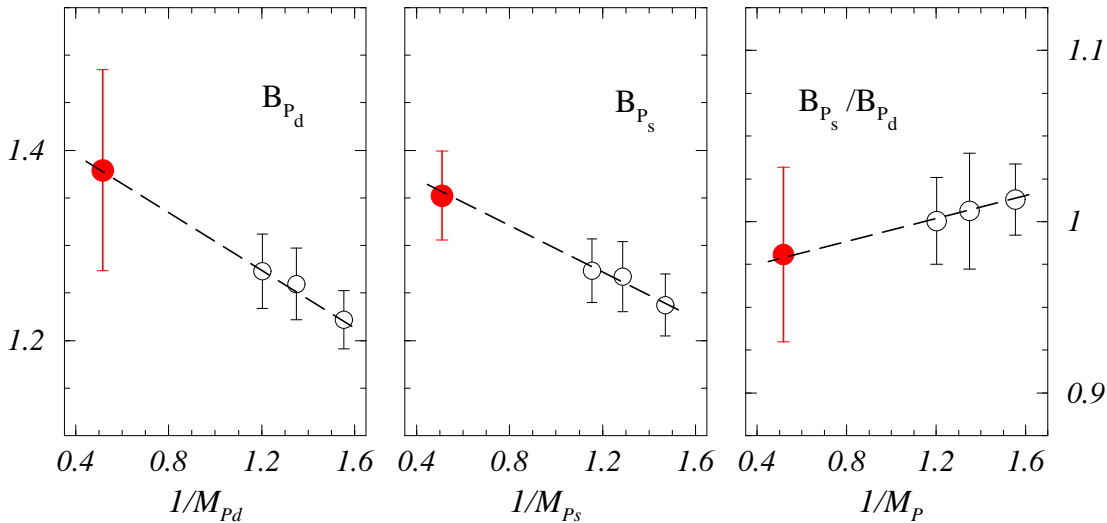} &    \\
\end{tabular}

\caption{\label{su3}{\sl Filled circles correspond to the B-meson sector. 
We see the $1/m_P$ negative slope for $\hat B_{B_d}$ and $\hat B_{B_s}$. Third picture 
shows the SU(3) breaking for $B$-parameter, which is clearly compatible 
with $1.0$ for all heavy-light mesons.}}
\end{center}
\end{figure}

In this section, we derive  the physical amplitudes from the
 $B$ parameters computed at fixed $m_Q$.
Before discussing the  extrapolation in the heavy-quark mass, let us 
summarize our results. We have computed the matrix elements
of the operator $Q(\mu)$ in the 
RI-MOM scheme at three different scales $\mu$. The dependence on light quark is
rather smooth, 
and we have easily performed the extrapolation/interpolation to the $d$ and $s$ 
 quarks. 
By using the NLO evolution, we have checked the  stability of our results. 
For each value  of the heavy-quark mass, from the $B$ parameter 
in the RI-MOM scheme we have computed 
${\hat B}_{P_q}$ by using the  result at $\mu \simeq 3.8$~GeV. 
This is the only step where perturbation theory comes in the game. 
At this point we dispose of the following values:
\bea
  \kappa_Q   &=& \{ \hspace*{8mm} 0.1250 \;\;\ \;, \hspace*{6mm} 0.1220 \;\;, 
  \hspace*{6mm}  0.1190\hspace*{7mm} \} 
\cr
&& \cr
  {\hat B}_{P_s} &=& \{ \  1.24(2)^{+0.00}_{-0.09}, \;\;  1.26(2)^{+0.00}_{-0.09}, \;\; 1.27(2)^{+0.00}_{-0.09}\ \}
\cr
&& \cr
 {\hat B}_{P_d} &=& \{ \ 1.22(3)^{+0.00}_{-0.09}, \;\;  1.26(4)^{+0.00}_{-0.09}, \;\; 1.27(4)^{+0.00}_{-0.09}\ \}
\eea
In order to extrapolate the above results ito the $b$ quark, 
 we rely on the heavy-quark symmetry as in the case of 
decay constants and fit the $\hat B$ parameters to the expression
\bea \label{scaling2}
\hat B_{P_q} \ = \ c_0^{(q)} \ \left( \ 1\ +\ {c_1^{(q)}\over M_P}\ \right) \; .
\eea
The result is shown in fig.~\ref{su3}, and the main
numbers are listed in tab.~\ref{tab6}. We stress  again that
we are not able to include terms of order $1/m_P^2$ in our fit, because we work
with three values of the heavy-quark mass only. Note, however, that our results for the slope 
$c_1^{(q)}$ are much smaller than the predictions of ref.~\cite{nrqcd},
where the heavy quark was treated non-relativistically.  We will comment on this
in the next section.
\begin{table}[h!]
\begin{center}
\begin{tabular}{|c|c|c|c|} 
\hline
{\phantom{\Large{l}}}\raisebox{+.2cm}{\phantom{\Large{j}}}
{ \hspace*{-4mm}${\hat B}_{ D_d}$}&{${\hat B}_{ B_d}$}  & ${\hat B}_{ B_s}$  & $({\hat B}_{ B_s}/{\hat B}_{B_d})$ \\    
{\phantom{\Large{l}}}\raisebox{-.4cm}{\phantom{\Large{j}}}
 $\mathsf 1.24(4)^{+0.00}_{-0.09}$ &  $\mathsf 1.38(11)^{+0.00}_{-0.09}$ &
  $\mathsf 1.35(5)^{+0.00}_{-0.08}$ & $\mathsf 0.98(5)$ \\ \hline  
 \multicolumn{3}{c}{ }   \\  
\end{tabular}
\vspace*{4mm}\\
\begin{tabular}{|c|c|} 
\hline
{\phantom{\Large{l}}}\raisebox{+.2cm}{\phantom{\Large{j}}}
{ \hspace*{-4mm}$c_0^{(q=d)}$}  & $c_1^{(q=d)}$  \\    
{\phantom{\Large{l}}}\raisebox{-.2cm}{\phantom{\Large{j}}}
 $1.46(2)$ &  $-0.28(16)\ {\rm GeV}$  \\ \hline
{\phantom{\Large{l}}}\raisebox{+.2cm}{\phantom{\Large{j}}}
  $c_0^{(q=s)}$  & $c_1^{(q=s)}$  \\    
{\phantom{\Large{l}}}\raisebox{-.2cm}{\phantom{\Large{j}}}
  $1.41(6)$ &  $-0.24(6)\ {\rm GeV}$   \\ \hline
\end{tabular}
\vspace*{.8cm}
\caption{\label{tab6}
{\sl Physical results for ${\hat B}_{P_q}$ parameters obtained in this work. In addition, we give the parameters $c_{0,1}^{(q)}$ from eq.~{\rm (\ref{scaling2})}.  }}
\end{center}
\end{table}
These results, when combined with those in  tab.~\ref{tab2}, give the following
final numbers
\bea
&& f_{B_d} \sqrt{{\hat B}_{B_d}} \ =\ 206(28)(7)\ \mev , \hspace*{8mm} 
 f_{B_s} \sqrt{{\hat B}_{B_s}} \ =\ 237(18)(8)\ \mev \ ;\cr
&& \cr
&&\xi \equiv {f_{B_s}\over f_{B_d}}\sqrt{{\hat B}_{B_s}\over  {\hat B}_{B_d}}\ =\ 1.16(7)\ , \hspace*{8mm} r_{sd} = \xi^2 \left({m_{B_s}^2 \over m_{B_d}^2}\right)^{\rm (exp.)} = 1.40(18) \;. 
\eea
In practice, for the reasons discussed in the introduction, we have extracted 
the value of $f_{B_d} \sqrt{\hat B_{B_d}}$, as well as the other quantities
appearing in the above equation, directly  
from the calculation of the mixing amplitude. 
\par
For completeness, and for comparison with other calculations, we also
give the $B$ parameter in the $\msbar$ scheme
\bea \hat B^{\msbar}_{B_d}(m_b)&=& 0.93(8)^{+0.0}_{-0.6} \ , \nonumber \\ 
\hat B^{\msbar}_{B_s}(m_b)&=& 0.92(3)^{+0.0}_{-0.6} \ . \eea 
\noindent

\vspace{1cm}

\section{Comparison with Other Calculations and Conclusions}
\label{sec:con}
In this section we compare our results with those obtained  with different
lattice calculations or with other methods.

\par On the lattice, two approaches have been  used so far to compute $B^0$--$\bar B^0$ 
mixing: the direct one (as done in this work) which 
extrapolates in the heavy-quark mass and the 
one where the heavy quark is treated with an effective theory. In the latter, 
either the 
Heavy Quark  Effective Theory (HQET)  or Non-Relativistic QCD (NRQCD) were used. 
There is a large number of 
lattice studies in which the HQET was invoked. Unfortunately, most of them were plagued 
by a mistake in the calculation of the  renormalization constant. Only recently, 
this has been corrected in ref.~\cite{hqet}, and the data of 
refs.~\cite{Gimenez,Ewing} have been reanalyzed~\footnote{This is 
the reason why we quote results from   these three papers only.}.
On the other hand, very recently, the NRQCD treatment of the heavy quark 
has been  employed to  compute $B_{B_q}$~\cite{nrqcd}. The authors also 
calculated  the relevant 
${\cal O}(1/m_P)$ corrections. The slope in $1/m_P$ that they observe 
for $B^{\msbar}_{B_d}(m_b)$ is roughly a factor of $3$ larger than ours 
(or  the one  by  ref.~\cite{ukqcd}). 
We  stress that there is a general argument which  shows
that the uncertainty  on the $1/m_P$ corrections can be easily as large as 
the corrections themselves unless perturbation theory on the leading
term is not pushed to very high orders~\cite{msreno}. This argument found
confirmation  by the explicit calculation of 
the ${\cal O}(\alpha^2_s)$ perturbative corrections to the heavy-quark mass
in the HQET~\cite{ms2loops}. Moreover, even the lowest order result of
ref.~\cite{nrqcd} in $1/m_P$ is  suspicious, since
they combine the renormalization constants computed in the HQET with the computation
of the matrix elements  of  NRQCD.  Until the renormalization
of the relevant operators will not be completed in a consistent 
theoretical framework,   the  results of ref.~\cite{nrqcd} should not be
used for comparisons with  other approaches. 
\par
\begin{figure}
\begin{center}
\begin{tabular}{@{\hspace{-0.7cm}}c c c}
&\epsfxsize12.0cm\epsffile{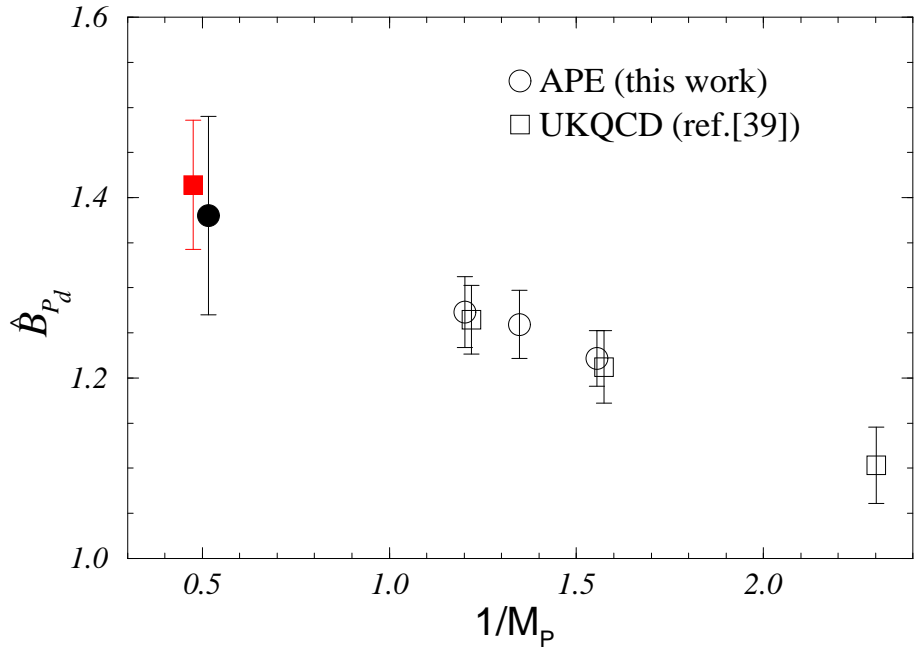} &  \\
\end{tabular}
\caption{\label{fig:comp}{\sl
Comparison of our results with those of ref.~{\rm \cite{ukqcd}}
as a function of the inverse heavy-meson mass in lattice units.}}
\end{center}
\end{figure}

In the direct approaches, a full relativistic treatment is given to both 
heavy and  light quarks~\cite{reviews}. In this case a major source
 of systematic error comes from the extrapolation
of the results to the physical $B$ mesons. 
 Recent results for $B^0$--$\bar B^0$ mixing 
can be found in refs.~\cite{ukqcd} and \cite{bbs}. In the second paper, an extrapolation
to  the continuum ($a\to 0$)  obtained using  non-improved Wilson 
fermions, has been attempted. They obtain a large central value for the 
ratio $r_{sd}$ ($r_{sd} = 1.76(10)^{+57}_{-42}$), albeit with large errors. 
The central value  is difficult to reconcile with the value of $\xi$ and
of the decay constants produced by the same authors with the same set of data. In the 
present work and the one by UKQCD~\cite{ukqcd},  improved fermions are used
but without extrapolation to the continuum. 
The differences between   the present study and that of~\cite{ukqcd}
 is that in our case the action is improved non-perturbatively 
whereas UKQCD uses the mean-field improved  action, 
and that we renormalize  non-perturbatively the relevant 
four-fermion operators.  The overall agreement is 
excellent.  This is true also at the values of the heavy-quark masses
were lattice measurements have been actually made as  shown in fig.~\ref{fig:comp}. Their extrapolated value 
has an error smaller than ours, probably because they also use a point corresponding to a small value of $m_P$,
for which the heavy-quark expansion may be questionable. 
 In  table~\ref{compare}, for  comparison, we list several recent 
results of the direct and the HQET approaches.

For completeness, let us mention that also QCD sum rules were  employed to compute 
$B_{B_d}$~\cite{pich}. Radiative corrections were also included in ref.~\cite{narison}, 
where the value for $B^{\msbar}_{B_d}(m_b)$ was found to be compatible 
with 1 (VSA value). The most recent QCD sum rule estimate has been given 
by Chernyak~\cite{chernyak}, who quotes a value lower than ours and closer to the 
results obtained from the HQET at lowest order in $1/m_P$:
 $B^{\msbar}_{B_d}(5.04\ \gev) = 0.82$, corresponding to 
${\hat B}_{B_d} = 1.26$. The effect of SU(3) breaking in $B_{B_s}/B_{B_d}$, 
was studied in the framework of the HQET, combined with chiral perturbation 
theory~\cite{grinstein}. Their result suggests that  SU(3) breaking 
for the $B_B$ parameter is practically negligible 
as confirmed by many lattice calculations. The approach has been extended to
quenched chiral perturbation 
theory~\cite{sharpe}, where a pessimistic estimate of quenching errors
($\sim 10$--$20 \%$ for the matrix element)  was quoted.
\begin{table}
\begin{center}
\begin{tabular}{|c|c|c|} 
\hline
{\phantom{\Large{l}}}\raisebox{+.5cm}{\phantom{\Large{j}}}
{ Reference} & $\;\;\;\;\; \;\;\;\;\;  {\hat B}_{B_d}\;\;\;\;\; \;\;\;\;\; 
$  & $\;\;\;\;\; \hat B_{ B_s}/\hat B_{B_d}\;\;\;\;\; $ \\    
&&\\  \hline  \hline
{\phantom{\Large{l}}}\raisebox{+.3cm}{\phantom{\Large{j}}}
{ This work}& $\mathsf 1.38(11)^{+.00}_{-.09}$ &  $\mathsf 0.98(5)$ \\    
&&\\ \hline 
 {\phantom{\Large{l}}}\raisebox{.3cm}{\phantom{\Large{j}}}
{\cite{ukqcd}}& $\mathsf 1.41(6)^{+.05}_{-.00}$ &  $\mathsf 0.98(3)$\\ 
&&\\ \hline  
 {\phantom{\Large{l}}}\raisebox{.3cm}{\phantom{\Large{j}}}
{\cite{bbs}}& $\mathsf 1.41(18)$ &  $\mathsf \simeq 1$\\   
&&\\ \hline  \hline  
 {\phantom{\Large{l}}}\raisebox{.3cm}{\phantom{\Large{j}}}
{\cite{hqet}+\cite{Gimenez}}& $\sl 1.29(8)(6)$ &  $\sl - $\\  
&&\\ \hline  
 {\phantom{\Large{l}}}\raisebox{.3cm}{\phantom{\Large{j}}}
{\cite{hqet}+\cite{Ewing}}& $\sl 1.26(6)(6)$ &  $\sl - $\\  
&&\\ \hline\hline  
\end{tabular}

\caption{\label{compare}{\sl Recent lattice results on ${\hat B}_{P_d}$ 
(at NLO accuracy). The upper part of the table refer to results obtained 
by extrapolation to the B-mesons, whereas in the lower part,  the 
HQET was used. The SU(3) breaking ratios are also 
listed (when available).}}
\end{center}
\end{table}
\section*{Acknowledgements} We warmly thank E.~Franco and D.~Lin for informative
discussions on the subject of this paper.  V.L. and G.M. acknowledge MURST for 
partial support. D.B. thanks INFN for support. 
V. G. has been supported by CICYT under 
Grant AEN-96-1718, by DGESIC under Grant PB97-1261 and by the Generalitat
Valenciana under Grant GV98-01-80. L. G. has been supported in part under 
DOE Grant DE-FG02-91ER40676.



\begin{thebibliography}{99}



\bibitem{UA1}  C.~Albajar {\em et al.}, UA1 Collaboration, Phys. Lett.
{\bf B186} (1987) 237, {\it ibid} 
{\bf B186} (1987) 247.

\bibitem{lusignoli}
M.~Lusignoli, L.~Maiani, G.~Martinelli and L.~Reina,
Nucl.\ Phys.\  {\bf B369} (1992) 139.

\bibitem{ali}
A.~Ali and D.~London,
Eur.\ Phys.\ J.\  {\bf C9} (1999) 687, 
{\tt [hep-ph/9903535]}.

\bibitem{enrico}
M.~Ciuchini, E.~Franco, L.~Giusti, V.~Lubicz and G.~Martinelli,
{\tt [hep-ph/9910236]}.


\bibitem{roudeau}
F.~Parodi, P.~Roudeau and A.~Stocchi, Nuovo Cim. {\bf A112} (1999) 833,
{\tt [hep-ex/9903063]}.

\bibitem {mele} S.~Mele, Phys.\ Rev.\  {\bf D59} (1999) 113011, 
{\tt [hep-ph/9810333]}.


\bibitem{babar} S.~Plaszcynsky and M.H.~Schune, {\tt [hep-ph/9911280]}.

\bibitem{barbieri} L.~Giusti, A.~Romanino and A.~Strumia,
Nucl.\ Phys.\  {\bf B550} (1999) 3,{\tt [hep-ph/9811386]};
R.~Barbieri, L.J.~Hall and A.~Romanino,
Nucl. Phys. {\bf 551} (1999) 93, {\tt [hep-ph/9812384]}.



\bibitem{buras}
G.~Buchalla, A.~J.~Buras and M.~E.~Lautenbacher,
Rev.\ Mod.\ Phys.\  {\bf 68} (1996) 1125, 
{\tt [hep-ph/9512380]}.

\bibitem{reviewbb} see for example
A.~J.~Buras, in ``Probing the
Standard Model of Particle Interactions'', 
F.David and R. Gupta, eds., 1998, Elsevier Science B.V.,
{\tt [hep-ph/9806471]}.


\bibitem{inami} T.~Inami and C.S.~Lim, Prog.~Theor.~Phys. {\bf 65} (1981) 297; {\em
ibid.} {\bf 65} (1981) 1772.


\bibitem{np} G.~Martinelli, C.~Pittori, C.T.~Sachrajda, M.~Testa and 
A.~Vladikas, Nucl. Phys.~{\bf B445}~(1995)~81, {\tt [hep-lat/9411010]}.

\bibitem{DELTAS=2} G.~Martinelli et al., Nucl.~Phys.~{\bf B445}~(1995)~81; \\
A.~Donini et al.,  Phys.~Lett.~{\bf B360}~(1996)~83; \\
M.~Crisafulli et al., Phys.~Lett.~{\bf B369}~(1996)~325; \\
L. Conti et al.,  Phys.~Lett.~{\bf B421}~(1998)~273.

\bibitem{bibbia}
A.~Donini, V.~Gimenez, G.~Martinelli, M.~Talevi and A.~Vladikas,
Eur.\ Phys.\ J.\  {\bf C10} (1999) 121, 
{\tt [hep-lat/9902030]}.



\bibitem{experiment}
M.~Daoudi, talk given at ``Heavy Flavors 8'',
Southampton, England, 25-29 Jul 1999, {\tt [hep-ex/9911027]}.


\bibitem{heavy}
D.~Becirevic {\em et al.}, Phys.\ Rev.\  {\bf D60} (1999) 074501, {\tt [hep-lat/9811003]}\ ;\\
A.~Abada {\em et al.}, presented at ``Lattice 99'', Pisa, Italy, 
29 June -- 3 July 1999, {\tt [hep-lat/9910021]}.



\bibitem{nir} Y.~Grossman, Y.~Nir and R.~Rattazzi,  
in 'Heavy Flavours II', eds. A.J. Buras and M. Lindner,
``Advanced Series on Directions in High Energy Physics'', 
World Scientific Publishing Co., Singapore, 
{\tt [hep-ph/9701231]}.


\bibitem{reviews}
L.~Lellouch, talk given at ``$34^{\acute{e}mes}$ Rencontres de Moriond", Les Arcs, France,
13-20 Mar 1999, {\tt [ hep-ph/9906497]};\\
J.~M.~Flynn and C.~T.~Sachrajda, in ``Heavy Flavours''
 (2nd ed.), 402-452, ed. by A.J. Buras and M. Linder 
 (World Scientific, Singapore), {\tt [hep-lat/9710057]}\ ;\\
H.~Wittig,
Int.\ J.\ Mod.\ Phys.\  {\bf A12} (1997) 4477,
{\tt [hep-lat/9705034]}.



\bibitem{csw} M.~L\"uscher, S.~Sint, R.~Som\-mer, H.~Wit\-tig,
 Nucl.~Phys.~{\bf B491} (1997) 344, {\tt [hep-lat/9611015]}.

\bibitem{light}
D.~Becirevic {\em et al.}, 
{\tt [ hep-lat/9809129]} and Phys.~Lett. {\bf B444} (1998) 401;  
{\tt [ hep-lat/9807046]}. 


\bibitem{lpmethod}
C.R. Allton, V. Gimenez, L. Giusti, F. Rapuano,
Nucl. Phys. {\bf B489} (1997)427, {\tt [hep-lat/9611021]}.


\bibitem{alpha} M.~L\"uscher, S.~Sint, R.~Som\-mer, P.~Weisz, U.~Wolff, Nucl.~Phys.~{\bf B491} (1997) 323, {\tt [hep-lat/9609035]}.


\bibitem{bA} S.~Sint, P.~Weisz, Nucl.~Phys.~{\bf B502} (1997) 251, {\tt [hep-lat/9704001]}.


\bibitem{lanl} T.~Bhattacharya, S.~Chandrasekharan, R.~Gupta, W.~Lee, S.~Sharpe,\ Phys.\ Lett. {\bf B461} (1999) 79, {\tt [hep-lat/9904011]}, {\it ibid} Nucl.\ Phys.\ Proc.\ Suppl.\  {\bf 73} (1999) 276, {\tt [hep-lat/9810018]}.


\bibitem{klm} G.P.~Lepage and P.B.~Mackenzie, Phys. Rev.{\bf D48} (1993) 2250,
{\tt [hep-lat/9209022]};\\
A.X.~El-Khadra, A.S.~Kronfeld and P.B.~Mackenzie,
Phys. Rev. {\bf D55} (1994) 3933, {\tt [hep-lat/9604004]}.



\bibitem{vittorio} D.~Becirevic, V.~Lubicz, G.~Martinelli, M.~Testa,
talk given at ``Lattice 99'', Pisa, Italy, 29 Jun - 3 Jul
1999, {\tt [hep-lat/9909082]}. 

\bibitem{ZA1} M.~L\"uscher, S.~Sint, R.~Som\-mer, H.~Wit\-tig, Nucl.~Phys.~{\bf B491} 
(1997) 344, {\tt [hep-lat/9611015]}.



\bibitem{braun}
V.~M.~Braun, talk given at ``Heavy Flavors 8'', Southampton, England, 25-29 Jul 1999, 
{\tt [hep-ph/9911206]}.



\bibitem{bernard}
C.~Bernard {\it et al.},
Phys.\ Rev.\ Lett.\  {\bf 81} (1998) 4812, 
{\tt [hep-ph/9806412]}.


\bibitem{milc}
C.~Bernard {\it et al.},  MILC Collaboration,
{\tt [hep-lat/9909121]}.

\bibitem{cppacs}
A.~Ali Khan {\it et al.},  CP-PACS Collaboration,
talks given at ``Lattice 99'', Pisa, Italy, 29 Jun - 3 Jul
1999, {\tt [hep-lat/9909052]}, and at ``EPS-HEP 99'', Tampere,
Finland, 15-21 Jul 1999, {\tt [hep-ph/9909398]} .

\bibitem{marti4f} G.~Martinelli,  Phys.~Lett.~{\bf 141B} (1984) 395.





\bibitem{ciuchini}
M.~Ciuchini {\em et al.},  Nucl.\ Phys.\  {\bf B523} (1998) 501, 
{\tt [hep-ph/9711402]}.


\bibitem{guido}
G.~Martinelli, talk given at ``Heavy Flavors 8'',
Southampton, England, 25-29 Jul 1999, to appear in the proceedings;
V.~Gimenez, L.~Giusti, G.~Martinelli and F.~Rapuano, LPT-Orsay/00-18,
{\tt [hep-ph/0002007]}.



\bibitem{nrqcd}
S.~Hashimoto, K.~I.~Ishikawa, H.~Matsufuru, T.~Onogi, S.~Tominaga and N.~Yamada,
Phys.\ Rev.\  {\bf D60} (1999) 094503
{\tt [hep-lat/9903002]};\\
N.~Yamada, S.~Hashimoto, K.~Ishikawa, H.~Matsufuru and T.~Onogi,
talk given at ``Lattice 99'', Pisa, Italy, 29 Jun - 3 Jul
1999, {\tt [hep-lat/9910006]}, {\it ibid} Nucl.\ Phys.\ Proc.\ Suppl.\  {\bf 73} (1999) 354, {\tt [hep-lat/9809156]}.


\bibitem{hqet}
V.~Gimenez and J.~Reyes,
Nucl.\ Phys.\  {\bf B545} (1999) 576,
{\tt [hep-lat/9806023]}.


\bibitem{Gimenez}
V.~Gimenez and G.~Martinelli,
Phys.\ Lett.\  {\bf B398} (1997) 135,
{\tt [hep-lat/9610024]}.


\bibitem{Ewing}
A.~K.~Ewing {\it et al.}, UKQCD Collaboration,
Phys.\ Rev.\  {\bf D54} (1996) 3526,
{\tt [hep-lat/9508030]}.


\bibitem{ukqcd}
L.~Lellouch and C.~J.~Lin, UKQCD collaboration, Nucl.\ Phys.\ Proc.\ Suppl.\  {\bf 73} (1999) 357, {\tt [hep-lat/9809018]}, and
talk given at ``Heavy Flavors 8'',
Southampton, England, 25-29 Jul 1999, {\tt [hep-ph/9912322]}.

\bibitem{msreno} G.~Martinelli and C.~Sachrajda,
 Nucl.~Phys.~{\bf B478} (1996) 660.
 
\bibitem{ms2loops}  G.~Martinelli and C.~Sachrajda,
Rome-1-98-1234, {\tt [hep-lat/9812001].}

\bibitem{sharpe}
S.~R.~Sharpe and Y.~Zhang, Phys.\ Rev.\  {\bf D53} (1996) 5125,
{\tt [hep-lat/9510037]};\\
S.~R.~Sharpe,
Nucl.\ Phys.\ Proc.\ Suppl.\  {\bf 53} (1997) 181,
{\tt [hep-lat/9609029]}.


\bibitem{bbs}
C.~Bernard, T.~Blum and A.~Soni,
Phys.\ Rev.\  {\bf D58} (1998) 014501, 
{\tt [hep-lat/9801039]}.



\bibitem{grinstein}
B.~Grinstein, E.~Jenkins, A.~V.~Manohar, M.~J.~Savage and M.~B.~Wise,
Nucl.\ Phys.\  {\bf B380} (1992) 369, 
{\tt [hep-ph/9204207]}.



\bibitem{pich}
A.~Pich, 
Phys.\ Lett.\  {\bf B206} (1988) 322,\\
A.~A.~Ovchinnikov and A.~A.~Pivovarov,
Sov.\ J.\ Nucl.\ Phys.\  {\bf 48} (1988) 120,\\
L.~J.~Reinders and S.~Yazaki,
Phys.\ Lett.\  {\bf B212} (1988) 245.


\bibitem{narison}
S.~Narison and A.~A.~Pivovarov,
Phys.\ Lett.\  {\bf B327} (1994) 341,
{\tt [hep-ph/9403225]}.


\bibitem{chernyak}
V.~Chernyak,
Nucl.\ Phys.\  {\bf B457} (1995) 96, 
{\tt [hep-ph/9503208]}.













\end{thebibliography}
\end{document}